\documentclass[a4paper,10pt]{article}
\usepackage[utf8]{inputenc}
\usepackage{natbib}
\usepackage[english]{babel}
\usepackage{array}
\usepackage{bm}
\usepackage{units}
\usepackage{amsmath}
\usepackage{amssymb}
\usepackage{graphicx}
\usepackage[usenames]{color}
\usepackage{psfrag}
\usepackage{float} 
\title{Equivalence Principle in Chameleon Models}

\author{
Lucila Kraiselburd$^{1}$\footnote{lkrai@fcaglp.fcaglp.unlp.edu.ar}, 
Susana Landau$^{2,3}$\footnote{ slandau@df.uba.ar}, 
Marcelo Salgado$^{4}$\footnote{marcelo@nucleares.unam.mx}
and Daniel Sudarsky$^{4}$\footnote{sudarsky@nucleares.unam.mx} \\ [5pt]
    $^1${\normalsize \it Grupo de Astrof\'{i}sica, Relatividad y Cosmolog\'{i}a, 
    Facultad de Ciencias Astron\'omicas y Geof\'{\i}sicas,} \\ [-2pt]
  {\normalsize \it Universidad Nacional de La PLata,} \\ [-2pt]
  {\normalsize \it Paseo del Bosque S/N (1900) La Plata, Argentina } \\
$^2${\normalsize \it Instituto de  F\'{i}sica de Buenos Aires, CONICET - 
Universidad de Buenos Aires,} \\ [-2pt] 
{\normalsize \it Ciudad Universitaria - Pab. 1, 1428 Buenos Aires, Argentina}\\ [-2pt]
$^3${\normalsize \it Departamento de  F\'{i}sica de Buenos Aires, FCEN,
Universidad de Buenos Aires,} \\ [-2pt] 
{\normalsize \it Ciudad Universitaria - Pab. 1, 1428 Buenos Aires, Argentina}\\ [-2pt]
$^4${\normalsize \it Instituto de Ciencias Nucleares, UNAM,} \\ [-2pt] 
{\normalsize \it A. Postal 70-543, M\'exico D.F. 04510, M\'exico.}
}

\begin{document}

\maketitle

\abstract{Most theories that predict time and/or space variation of fundamental constants also predict violations of the Weak Equivalence Principle (WEP). Khoury and Weltmann proposed the chameleon model in 2004 and claimed that this model avoids experimental bounds on WEP. We present a contrasting  view  based  on an  approximate calculation of the two body problem for the chameleon field and show that the force depends on the test body composition. Furthermore, we compare the prediction of the force on a test body with E\"{o}tv\"{o}s type experiments and find that the chameleon field effect cannot account  for current bounds. 

}
\section{Introduction}
\label{Intro}

Most theories that predict variation of fundamental constants, also predict  violations of the Weak Equivalence Principle (WEP) \citet{Bekenstein82,Barrow02,Olive02,DP94}, \citet{Palma03}. The reason for this is that the mass of a body is made of many contributions related to various interaction energies (strong, weak, electromagnetic). Therefore any theory in which the local coupling constants become effectively  spatially dependent through their  direct dependence on  an light scalar field will entail  some non-universality in the free-fall acceleration of bodies embedded in an external gravitational field. On the other hand, WEP is strongly constrained by E\"{o}tv\"{o}s type experiments; latest results give  $\frac{\Delta a}{a} \simeq 10^{-14}$. However, some schemes   claim to be able to avoid this problem  based on proposals such as the  chameleon models and the Dilaton-Matter-gravity model with strong coupling \citet{Damour02}. Chameleon models where introduced by \citet{KW04} and further developments were performed by several authors \citet{Brax04,MS07}.
Khoury and Weltman, 2004 have shown that the parameters of the chameleon model were constrained by experimental bounds of the WEP. On the other hand, Mota \& Shaw, 2007 claim that while the linear and quasi-linear solution seems to predict violation of WEP, the non-linear solution avoids that prediction  at the particle level. According to these authors, the reason for this lies in that  the non-linear effects   become relevant only on a small region near the body's  surface which has  been denominated ``{\it the  thin shell}". Based  on such  analysis, it has  been   argued that the  chameleon  field  does  not  depend on the composition of the falling  body or in general on the  interaction between the  matter   and chameleon fields. In this paper, we perform an alternative calculation of the chameleon mediated force on a free falling body exerted by a large body like a mountain or the Earth. First, we propose an approximate solution of the two body problem  and show that the force on a test body depends on its composition. Furthermore we compare the prediction for the differential acceleration between two test   objects of different composition,  subject to the  Earth's  or Suns  attraction,   with the corresponding  observational bounds  extracted  from E\"{o}tv\"{o}s type  experiments.
\section{The Model}

\label{modelo}
Let us first briefly review the chameleon model. The theory is  characterized  by  the general action: 
\begin{eqnarray}
  S &=& \int d^4x \sqrt{-g} \left[ \frac{M_{pl}}{2} R - (\partial \Phi)^2 - V(\Phi)\right] \nonumber \\
& & - \int d^4x L_m \left(\Psi_m^{(i)}, g_{\mu \nu}^{(i)}\right) 
\end{eqnarray}

where each matter field $\Psi^{(i)}$ couples to a metric $g_{\mu\nu}^{(i)}$ related to the Einstein-frame metric $g_{\mu \nu}$ by a conformal factor: $g_{\mu\nu}^{(i)} = \exp{\frac{2 \beta_i \Phi}{M_{pl}}} g_{\mu\nu}$, $M_{pl}$ is the Planck mass  and  the potential is assumed to be 
$V(\Phi)= M^{4+{\it n}} \Phi^{\it -n}$ where $M$ is a constant. {In order to exhibit the chameleon effect we shall assume $\beta_i= \beta$, although it is precisely the 
fact that each species of particle couples differently to the chameleon field the reason behind the potential violation of the WEP, and 
so the {\it thin shell effect} has  been  considered   as  the  crucial  mechanism required to suppress such violations.
The dynamics of the chameleon depends crucially on its effective potential according to the following equation: 
%The key ingredient: The non-linear effects are only relevant for a small region near to the body surface named thin shell

\begin{eqnarray}
%\begin{equation}
\label{mov}
\Box \Phi &=& \frac{\partial V_{eff}}{\partial \Phi} \\
V_{eff} &=& V(\Phi) - T \exp(\frac{\beta \Phi}{M_{pl}}) \nonumber
%\end{equation}
\end{eqnarray}
where T is  the  trace of the  energy-momentum tensor of the matter  occupying the region under consideration.

In order to compute the force on a test body that is free-falling in the presence of a larger body, we have to solve the above equation for the case of two bodies. In this paper, we present an approximate solution for the chameleon field in the presence of two spherical bodies. We expand the most general solution in terms of complete sets of solutions in three regions:i) Inside the large body, ii) Inside the test body and iii) Outside both bodies. For region i) and iii) we keep the dominant term: the one-body solution. Inside the free falling body we seek the most general solution for the two body problem, and impose continuity at the border.

 For the benefit of the reader we reproduce the solution of Eq.~(\ref{mov}) for a spherically-symmetric body of radius $R$ and density $\rho_{in}$ immersed in an external medium of density $\rho_{out}$ (for details see \citet{KW04}) being $T_{\rm in (out)}=-\rho_{\rm in (out)}+3P_{\rm in (out)}$. In this case we must consider 2 regions and $3P\to 0$:

\begin{equation*}
\quad \rho=
\begin{cases}
\rho_{\rm in} \hspace{1cm} r \leq R \\
\\
\rho_{\rm out} \hspace{1cm} r > R
\end{cases}
\nonumber
\end{equation*}
The  next step is to   make   suitable  expansions of the  effective potential  about the  corresponding minimum  both  for the outside  and inside  regions:

\begin{eqnarray}
V_{eff}^{\rm in,out}(\Phi) &\simeq& V_{eff}^{\rm in,out}(\Phi_{\rm min}^{\rm in,out}) \nonumber \\ & &+\frac{1}{2} \partial_{\Phi \Phi} V_{eff}^{\rm in,out}(\Phi_{\rm min}^{\rm in,out})(\Phi - \Phi_{\rm min}^{\rm in,out}) \nonumber
\end{eqnarray}
where the superscript ${\rm in (out) }$ refers to inside (outside) of the large body.
We define:

\begin{equation}
m_{eff}^{ 2 \rm in,out}(\Phi_{\rm min}^{\rm in, out}, \beta, T_{\rm in,out})=  \partial_{\Phi \Phi} V_{eff}^{\rm in,out}(\Phi_{\rm min}^{\rm in,out}) \nonumber
\end{equation}
Thus  we  are  lead  to solve the following equations:

\begin{eqnarray}
\frac{1}{r}\partial r(r^2 \partial r \Phi^{\rm in}) = m_{eff}^{2 \rm in} (\Phi^{\rm in} - \Phi_{\rm min}^{\rm in})\nonumber \\
\frac{1}{r}\partial r(r^2 \partial r \Phi^{\rm out}) = m_{eff}^{2 \rm out} (\Phi^{\rm out} - \Phi_{\rm min}^{\rm out})\nonumber 
\end{eqnarray}

with the border conditions:

\begin{eqnarray}
\Phi^{\rm in}(r=0)&=&\Phi_0 \nonumber \\
\partial r \Phi^{\rm in} (r=0) &=&0 \nonumber \\
\Phi^{\rm out}(r \rightarrow \infty) &=& \Phi_{\rm min}^{\rm out}= \Phi_\infty \nonumber 
\end{eqnarray}

and the condition for both solutions ($\rm in,out$) to match is:
\begin{eqnarray}
\Phi^{\rm in} (r=R) &=& \Phi^{\rm out} (r=R) \nonumber \\
\partial r \Phi^{\rm in} (r=R)&=& \partial r \Phi^{\rm out} (r=R) \nonumber
\end{eqnarray}

The solution for the one body problem for $r \le R$ is:
\begin{equation}
\Phi^{\rm in}(r) = \frac{(\Phi_0 - \Phi_c) \sinh(m_{eff}^{\rm in} r)}{m_{eff}^{\rm in} r} + \Phi_c \nonumber 
\end{equation}
where $\Phi_c = \Phi_{\rm min}^{\rm in}$ and 
\begin{equation}
\Phi_0= \Phi_c + 
\frac{(\Phi_\infty - \Phi_c) \left[ 1 + m_{eff}^{\rm out} R \right]}{\frac{m_{eff}^{\rm out} R}{x} \sinh(x) + \cosh(x)} \nonumber 
\end{equation}
with $x = m_{eff}^{in} R$. For $r \ge R $ we get:
\begin{equation}
\Phi^{\rm out}(r)= C \frac{\exp{(-m_{eff}^{\rm out} r)}}{r} + \Phi_{\infty} \nonumber 
\end{equation}
where 
\begin{equation}
C = \frac{R (\Phi_c - \Phi_\infty) \left[ \cosh(x) - \frac{\sinh(x)}{x}\right] \exp{(m_{eff}^{out} R)}}{\frac{m_{eff}^{\rm out} R}{x} \sinh(x) + \cosh(x)}
\nonumber 
\end{equation}
Now we take the thin-shell approximation  (i.e. $1 \ll m_{eff}^{in} R$) and for simplicity assume $m_{eff}^{\rm out} R \ll 1$, 
and obtain approximate expressions for the field inside and outside the large body respectively:

\begin{eqnarray}
\Phi^{\rm in}(r) &\approx& \frac{2 (\Phi_\infty - \Phi_c) \exp{[-R m_{eff}^{\rm in}]} \sinh(m_{eff}^{\rm in} r)}{m_{eff}^{\rm in} r} \nonumber \\
 & & + \Phi_c \nonumber 
\end{eqnarray} 
\begin{eqnarray}
\Phi^{\rm out}(r)&\approx & R (\Phi_c - \Phi_{\infty}) \exp{(m_{eff}^{\rm out} R)} \times  \nonumber \\
& & \frac{\exp{(-m_{eff}^{\rm out} r)}}{r} + \Phi_{\infty} \nonumber 
\end{eqnarray}
 Notice that due to the exponential factor, the $r$ dependence in $\Phi^{\rm in}(r)$ is quite suppressed well inside the body
(where $\Phi^{\rm in}(r) \approx \Phi_c$), and only within a {\it thin shell} near the surface of the body the field grows to match the exterior solution 
$\Phi^{\rm out}(r)$.

%\begin{figure}
%\begin{center}
%\includegraphics[scale=0.22]{figura2.eps}
%\caption{ Regi\'{o}n de $95\%$ de confianza para la probabilidad de 1 par\'{a}metro}
%\end{center}
%\caption{WMAP collaboration - Planck collaboration}
%\label{figura2}
%\end{figure}
\begin{figure}[h]
\begin{center}
\includegraphics[width=4.5cm,height=6cm]{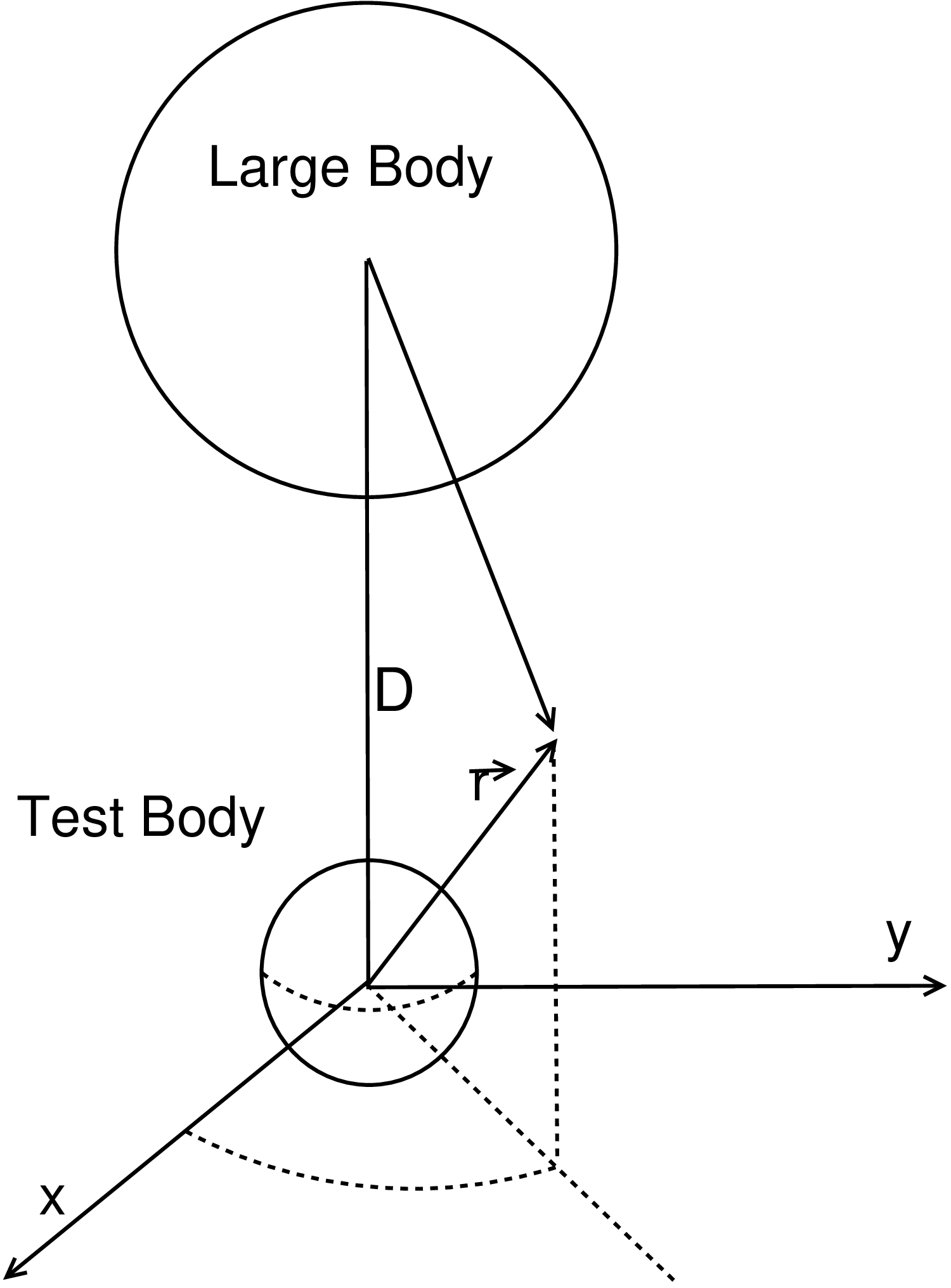}
\caption{\footnotesize Two body problem.}
\label{figura1}
\end{center}
\end{figure}
Figure \ref{figura1} depicts the two body problem considered below. In order to solve it, 
first we expand the most general solution for the two body problem in complete sets of solutions inside and outside the test body:
\begin{equation}
\quad \Phi(r,\phi,\theta)=
\begin{cases}
  \sum_{lm} C_{lm}^{in} i_l(\mu r) Y_{lm}(\theta,\phi) \hspace{0.25cm} r \le R_{2} \\
\sum_{lm} C_{lm}^{out} k_l(\tilde\mu r) Y_{lm}(\theta,\phi)\hspace{0.25cm}r > R_{2} 
\end{cases}
\label{phi}
\end{equation}
where $i_l(\mu r)$ and $k_l(\tilde\mu r)$ are the  spherical modified Bessel functions;   $\tilde{\mu}=m_{eff}^{out}$ and $\mu=m_{eff}^{\rm test\hspace{0.1cm} body}$; and $R_2=R_{\rm test\hspace{0.1cm} body}$.

Now we assume that the chameleon field outside the test body  can be taken  as the one body problem solution:
%\begin{eqnarray}
%  \Phi^{out}(\rho) &=& \tilde\mu C k_0 (\tilde \mu \rho) + \Phi_\infty \nonumber \\
%&=& C \frac{\exp{(-\tilde \mu \rho)}}{\rho}+ \Phi_\infty \nonumber
%\end{eqnarray}
\begin{equation}
\Phi^{\rm out}(\rho)=C \frac{\exp{(-\tilde \mu \rho)}}{\rho}+ \Phi_\infty \nonumber
\end{equation}
where $\vec \rho = \vec r - D \hat z$. 

 We shall use the following identity to rewrite $\Phi^{\rm out}(\rho)$ in terms of the coordinate system centered in the middle of the test body:
\begin{eqnarray}
\frac{\exp{(- \tilde\mu |\vec r_2 - \vec r_1|)}}{4 \pi |\vec r_2 - \vec r_1|}& =& \tilde{\mu} \sum_{l=0}^\infty i_l(\tilde{\mu} r_2) k_l(\tilde{\mu} r_1 ) \times  \nonumber \\ & & \sum_{m=-l}^l Y_{lm}(\theta_2,\phi_2) Y_{lm}^*(\theta_1,\phi_1) \nonumber
\end{eqnarray}

where $r_2 < r_1$. For the problem at hand  $\vec r_1 = D \hat z$, $\theta_1=0$, $\vec r_2 = \vec r$.The problem we are considering is symmetric respect to rotations around the z axis, and therefore, only the term with $m=0$ contributes to the solution. Now we match both solutions as follows:

\begin{equation}
\Phi_{\rm in}(R_{2}) = \Phi_{\rm out}(R_{2})
\nonumber
\end{equation}

and find:

%\begin{equation}
%C_{l0} = \frac{2(\Phi_\infty - \Phi_0) + C \tilde{\mu} i_l(\tilde \mu R_2) k_l(\tilde{\mu} D)}{\sqrt{4 \pi (2 l + 1) i_l(\mu R_2)} } \nonumber
%\label{cl}
%\end{equation}

\begin{eqnarray}
C_{l0} &=& \left[(\Phi_\infty - \Phi^{\rm 2 min}_{\rm in})\delta_{l0} + C \tilde{\mu} i_l(\tilde \mu R_2) k_l(\tilde{\mu} D)\right] \times \nonumber \\
&&\frac{\sqrt{4 \pi (2 l + 1)}} {i_l(\mu R_2)}  
\label{cl}
\end{eqnarray}
where $\Phi^{\rm 2 min}_{\rm in}$ is now the value of $\Phi$ that makes the effective potential of the two body problem reach its minimum  inside the test body.

\section{Force on a free falling body}

In this section we calculate the force on a free falling test body using the approximate solution for the chameleon field we have found in section \ref{modelo}. 

%\begin{eqnarray}
%F_z &=&  \int_V d^3x T  \frac{\partial \Phi}{\partial z}d^3x \nonumber \\
%& = &  (\rho - 3 P)  \int _V d^3x \frac{\partial  \exp[{\frac{\beta \Phi}{M_{pl}}}]}{\partial z}\nonumber \\
%& = &  \frac{2 \pi \beta}{M_{Pl}} T \times \nonumber \\
%& & \int_0^{R_2} d\rho \left[ \exp[\frac{\beta }{M_{pl}} \Phi(Y+\rho)] - \exp[\beta \Phi(Y-\rho)] \right] \nonumber \\
%& \simeq &  \frac{2 \pi \beta}{M_{Pl}} T \int_0^{R_2} d\rho[ \Phi(Y+\rho) -  \Phi(Y-\rho)],
%\end{eqnarray}
\begin{eqnarray}
F_{z} &=& - \int_{V} d^3x T \frac{\beta}{M_{pl}} e^{ \frac{\beta}{M_{pl}}\Phi} \vec{\nabla}\Phi \nonumber \\
& = &  - T  \int _V d^3x \frac{\partial  \exp[{\frac{\beta \Phi}{M_{pl}}}]}{\partial z}\nonumber \\
%&&  \int_V d^3x T  \frac{\partial \Phi}{\partial z}d^3x \nonumber \\
%& = &   (\rho - 3 P)  \int _V d^3x \frac{\partial  \exp[{\frac{\beta \Phi}{M_{pl}}}]}{\partial z}\nonumber \\
& = &  -2 \pi T  \int_0^{R_2}\rho d\rho \left[ e^{[\frac{\beta }{M_{pl}} \Phi(Y+\rho)]} - e^{[\frac{\beta }{M_{pl}} \Phi(Y-\rho)]} \right] \nonumber \\
& \simeq &   -\frac{2 \pi T\beta}{M_{Pl}}\int_0^{R_2}\rho d\rho[\Phi(Y+\rho)-\Phi(Y-\rho)] 
\end{eqnarray}
being $Y=D+R_2$. Now we evaluate the field using Eqs. (\ref{phi}) and (\ref{cl}) considering only the term with $l=0$:

\begin{eqnarray}
F_z & \simeq &\frac{4 \pi T \beta}{M_{pl}} \left[\left(\frac{C \exp{(-\tilde{\mu} D)}}{\tilde{\mu} R_2 D}-\left(\Phi_{\rm in}^{\rm 2 min}-\Phi_\infty\right) \right)\right. \times \nonumber  \\
&& \left. \frac{R_2\sinh(\mu (D+R_2))}{\mu } \right] \label{fuerza}
\end{eqnarray}
From the expression above, it follows that the force has an important dependence with the distance between the test body and the large body through the term $\sinh(\mu (D + R_2))$. 
Now we can compare the predictions of the chameleon model using Eq.~(\ref{fuerza}) with experimental bounds on the differential acceleration of two bodies of different composition: $\eta=2\frac{|a_1-a_2|}{|a_1+a_2|}$. For the case of the test body we must consider  $T=-\rho+3P$ since in this case $\rho \simeq 10^{11}{\rm cm}^{-4}$ and  $3P \simeq 10^{8} {\rm cm}^{-4}$. Table \ref{abun} shows the values of the coupling of the chameleon to matter  $\beta$ for each value of the parameter of the potential ${\it n}$ such that the large body satisfies the thin shell condition. For the values shown in table \ref{abun} we fixed $M=1000 [{\rm cm}^{-1}]$ (the other free parameter of the chameleon potential), but we have also calculated  $\beta$ for lower values of $M=1,10,100 [{\rm cm}^{-1}]$; for these cases the minimum value for $\beta$ to satisfy the thin shell condition are larger than the values shown in table \ref{abun} and thus we the corresponding values of $\eta$ are larger than current experimental bounds. . Table \ref{abun3} shows the value of $\eta$ for pairs of test bodies with different composition free-falling in the gravitational field of the earth; the experimental bounds for the same test bodies  are shown in Table \ref{abun1}. We have also calculated $\eta$ for experiments studying free falling bodies towards the sun; in this case the difference between the accelerations of the test bodies is so large that $\frac{a_1}{|a_1+a_2|} \simeq 1$ and $\frac{a_2}{|a_1+a_2|} \simeq 0$ and therefore in all these  situations  we  can  use $\eta = 2$. We obtain similar results for the bodies falling through the earth (see table \ref{abun3}).These results suggest that the chameleon model might be ruled out by E\"{o}tv\"{o}s type experiments. 
Since we are working with an approximate solution of the chameleon field equation, we avoid reaching to stronger conclusions. Work in progress includes finding an exact solution of the two body problem to verify results shown in this paper.

\begin{table}
\caption{Values of {\it n} (parameter of the chameleon potential) and $\beta$ intervals (coupling of the chameleon to matter)  such that the Sun and the Earth satisfy the thin shell condition ($M=1000.[{\rm cm}^{-1}]$).}
\label{abun}
\begin{center}
\begin{tabular}{lcc}
\hline
\\
{\it n} & $\beta$ for the Sun & $\beta$ for Earth  \\
\hline
\\
1  &$ (10^{-13},10^{-1}) $ &$ (10^{-11},10^{-1}) $  \\
2  &$ (10^{-16},10^{-1}) $ &$ (10^{-13},10^{-1}) $ \\
3  &$ (10^{-17},10^{-1}) $ &$ (10^{-14},10^{-1}) $ \\
4  &$ (10^{-19},10^{-1}) $ &$ (10^{-15},10^{-1}) $  \\
5  &$ (10^{-20},10^{-1}) $ &$ (10^{-16},10^{-1}) $  \\

\hline
\end{tabular}
\end{center}
\end{table}

\begin{table}
  \caption{Results of the E\"{o}tv\"{o}s experiments (for original references see \citep{KV12}). Columns $1$ and $2$ show the composition of the bodies that are free-falling, column $3$ indicates whether the experiment measures the free fall to the Earth or Sun, column $4$ shows the experimental bound on $\eta$.  }
\label{abun1}
\begin{center}
\begin{tabular}{lcccc}
\hline
\\
Body 1 & Body 2 & source & $(\eta\pm\sigma_{\eta})\times10^{11}$  \\
\hline
\\
    Al & Au & Sun & 1.3 $\pm$ 1.5 \\
    Al & Pt & Sun & 0.03 $\pm$ 0.04 \\
    Si/Al & Cu & Sun & 0.51 $\pm$ 0.67 \\
    Moon-Like & Earth-Like & Sun &  0.005 $\pm$ 0.089 \\ 
    Be & Ti & Sun & - 0.031 $\pm$ 0.045 \\
    Be & Al & Sun & 0.0 $\pm$ 0.042 \\
    Be & Ti & Earth &  0.003 $\pm$ 0.018 \\
    Be & Al & Earth & -0.015 $\pm$ 0.015 \\
    Be & Al & Earth & -0.02 $\pm$ 0.28 \\
    Be & Cu & Earth & -0.19 $\pm$ 0.25 \\
\hline
\end{tabular}
\end{center}
\end{table}

\begin{table*}
\caption{Predictions for the differential acceleration of free-falling bodies in the gravitational field of the earth within the approximation considered in this paper. Columns $1$ and $2$ show the composition of the bodies that are free falling, column 3 shows the value of {\rm n} (free parameter of the chameleon potential), column 4 shows the value of $\beta$ (coupling of the chameleon to matter) and column 5 shows the value of the differential acceleration $\eta$.
}
\label{abun3}
\begin{center}
\begin{tabular}{lcccc}
\hline
\\
Test body 1 & Test body 2 & {\it n} & $\beta$ & $\eta_{chameleon}$  \\
\hline
\\
    Be & Ti & 1 & $(10^{-10},10^{-5})$ & 2.000 \\
       &   &  1 & $10^{-11}$ & 1.939\\
      &  & 2 & $(10^{-12},10^{-6})$ & 2.000\\
      &  & 2 & $10^{-13}$ & 1.860 \\
     &  & 3 & $(10^{-13},10^{-7})$ & 2.000 \\
     &  & 3 & $10^{-14}$ & 1.944 \\
     &  & 4 & $(10^{-14},10^{-8})$ & 2.000 \\
     &  & 4 & $10^{-15}$ & 1.850 \\
     &  & 5 & $(10^{-14},10^{-8})$ & 2.000 \\
     &  & 5 & $10^{-15}$ & 1.996 \\
     &  & 5 & $10^{-16}$ & 1.498 \\
\hline
\\
    Be & Al & 1 & $(10^{-10},10^{-5})$ & 2.000 \\
       &  & 1 & $10^{-11}$ & 1.231\\
      &  & 2 & $(10^{-11},10^{-6})$ & 2.000 \\
      &  & 2 & $10^{-12}$ & 1.982 \\
      &  & 2 & $10^{-13}$ & 1.059 \\
     &  & 3 & $(10^{-12},10^{-7})$ & 2.000 \\
     &  & 3 & $10^{-13}$ & 1.992 \\
     &  & 3 & $10^{-14}$ & 1.287 \\
     &  & 4 & $(10^{-13},10^{-8})$ & 2.000 \\
     &  & 4 & $10^{-14}$ & 1.959 \\
     &  & 4 & $10^{-15}$ & 1.060 \\
     &  & 5 & $(10^{-14},10^{-8})$ & 2.000 \\
     &  & 5 & $10^{-15}$ & 1.716 \\
     &  & 5 & $10^{-16}$ & 0.686 \\
%\hline
\hline
\\
    Be & Cu & 1 & $(10^{-11},10^{-5})$ & 2.000 \\
      &  & 2 & $(10^{-12},10^{-6})$ & 2.000 \\
      &  & 2 & $10^{-13}$ & 1.998 \\
     &  & 3 & $(10^{-14},10^{-7})$ & 2.000 \\
     &  & 4 & $(10^{-14},10^{-8})$ & 2.000 \\
     &  & 4 & $10^{-15}$ & 1.997 \\
     &  & 5 & $(10^{-15},10^{-8})$ & 2.000 \\
     &  & 5 & $10^{-16}$ & 1.994 \\

%\hline
%    Si/Al & Cu & 1 & $(10^{-13},10^{-1})$ & 2.000 \\
%      &  & 2 & $(10^{-16},10^{-1})$ & 2.000 \\
%     &  & 3 & $(10^{-17},10^{-1})$ & 2.000 \\
%     &  & 4 & $(10^{-19},10^{-1})$ & 2.000 \\
%     &  & 5 & $(10^{-20},10^{-1})$ & 2.000 \\
%\hline
%    Moon-Like & Earth-Like & 1 & $(10^{-13},10^{-1})$ & 2.000 \\
%      &  & 2 & $(10^{-16},10^{-1})$ & 2.000 \\
%     &  & 3 & $(10^{-17},10^{-1})$ & 2.000 \\
%     &  & 4 & $(10^{-19},10^{-1})$ & 2.000 \\
%     &  & 5 & $(10^{-19},10^{-1})$ & 2.000 \\
%     &  & 5 & $10^{-20}$ & 1.999 \\
%\hline
%    Be & Ti & 1 & $(10^{-13},10^{-1})$ & 2.000 \\
%      &  & 2 & $(10^{-16},10^{-1})$ & 2.000 \\
%     &  & 3 & $(10^{-17},10^{-1})$ & 2.000 \\
%     &  & 4 & $(10^{-19},10^{-1})$ & 2.000 \\
%     &  & 5 & $(10^{-20},10^{-1})$ & 2.000 \\
%\hline
%    Be & Al & 1 & $(10^{-13},10^{-1})$ & 2.000 \\
%      &  & 2 & $(10^{-16},10^{-1})$ & 2.000 \\
%     &  & 3 & $(10^{-17},10^{-1})$ & 2.000 \\
%     &  & 4 & $(10^{-19},10^{-1})$ & 2.000 \\
%     &  & 5 & $(10^{-20},10^{-1})$ & 2.000 \\
\hline
\end{tabular}
\end{center}
\end{table*}
%\begin{table}
%\caption{Results of the E\"{o}tv\"{o}s experiments \citep{KV12}.}
%\label{abun1}
%\begin{center}
%\begin{tabular}{lcccc}
%\hline
%\\
%Test body 1 & Test body 2 & source & $(\eta\pm\sigma_{\eta})\times10^{11}$  \\
%\hline
%\\
%    Al & Au & Sun & 1.3 $\pm$ 1.5 \\
%    Al & Pt & Sun & 0.03 $\pm$ 0.04 \\
%    Si/Al & Cu & Sun & 0.51 $\pm$ 0.67 \\
%    Moon-Like & Earth-Like & Sun &  0.005 $\pm$ 0.089 \\ 
%    Be & Ti & Sun & - 0.031 $\pm$ 0.045 \\
%    Be & Al & Sun & 0.0 $\pm$ 0.042 \\
%    Be & Ti & Earth &  0.003 $\pm$ 0.018 \\
%    Be & Al & Earth & -0.015 $\pm$ 0.015 \\
%    Be & Al & Earth & -0.02 $\pm$ 0.28 \\
%    Be & Cu & Earth & -0.19 $\pm$ 0.25 \\
%\hline
%\end{tabular}
%\end{center}
%\end{table}

\section*{Acknowledgements}
L.K. and S.L. are supported  by PIP 0152/10 CONICET and grant G119 UNLP.  M.S. is partially supported by PAPIIT grant IN107113. D.S. is supported in part by the CONACYT grant No. 101712. and by UNAM-PAPIIT grant IN107412.

%\end{acknowledgements}

%\bibliographystyle{aa}

\end{document}